\documentclass[showpacs,preprint,aps]{revtex4}

\usepackage{stmaryrd}
\usepackage{amstext}
\usepackage{amsfonts}
\usepackage{amsmath}
\usepackage{amssymb}
\usepackage{epsfig}
\usepackage{graphicx}%
\ifx\pdfoutput\relax\let\pdfoutput=\undefined\fi
\newcount\msipdfoutput
\ifx\pdfoutput\undefined\else
\ifcase\pdfoutput\else
\ifx\paperwidth\undefined\else
\ifdim\paperheight=0pt\relax\else\pdfpageheight\paperheight\fi
\ifdim\paperwidth=0pt\relax\else\pdfpagewidth\paperwidth\fi
\fi\fi\fi
\begin{document}
\title{A self-guide method to search maximal Bell violations for unknown quantum states}

\author{Li-Kai Yang}
\affiliation{Key Laboratory of Quantum Information, University of
Science and Technology of China, CAS, Hefei, 230026, China}
\affiliation{Synergetic Innovation Center of Quantum Information and Quantum Physics, University of Science and Technology of China, Hefei, Anhui 230026, China}

\author{Geng Chen$\footnote{email:chengeng@ustc.edu.cn}$}
\affiliation{Key Laboratory of Quantum Information, University of
Science and Technology of China, CAS, Hefei, 230026, China}
\affiliation{Synergetic Innovation Center of Quantum Information and Quantum Physics, University of Science and Technology of China, Hefei, Anhui 230026, China}

\author{Wen-Hao Zhang}
\affiliation{Key Laboratory of Quantum Information, University of
Science and Technology of China, CAS, Hefei, 230026, China}
\affiliation{Synergetic Innovation Center of Quantum Information and Quantum Physics, University of Science and Technology of China, Hefei, Anhui 230026, China}

\author{Xi-Xiang Peng}
\affiliation{Key Laboratory of Quantum Information, University of
Science and Technology of China, CAS, Hefei, 230026, China}
\affiliation{Synergetic Innovation Center of Quantum Information and Quantum Physics, University of Science and Technology of China, Hefei, Anhui 230026, China}

\author{Shang Yu}
\affiliation{Key Laboratory of Quantum Information, University of
Science and Technology of China, CAS, Hefei, 230026, China}
\affiliation{Synergetic Innovation Center of Quantum Information and Quantum Physics, University of Science and Technology of China, Hefei, Anhui 230026, China}

\author{Chuan-Feng Li$\footnote{email:cfli@ustc.edu.cn}$}
\affiliation{Key Laboratory of Quantum Information, University of
Science and Technology of China, CAS, Hefei, 230026, China}
\affiliation{Synergetic Innovation Center of Quantum Information and Quantum Physics, University of Science and Technology of China, Hefei, Anhui 230026, China}

\author{Guang-Can Guo}
\affiliation{Key Laboratory of Quantum Information, University of
Science and Technology of China, CAS, Hefei, 230026, China}
\affiliation{Synergetic Innovation Center of Quantum Information and Quantum Physics, University of Science and Technology of China, Hefei, Anhui 230026, China}

\begin{abstract}
In recent decades, a great variety of researches and applications concerning Bell nonlocality have been developed with the advent of quantum information science. Providing that Bell nonlocality can be revealed by the violation of a family of Bell inequalities, finding maximal Bell violation (MBV) for unknown quantum states becomes an important and inevitable task during Bell experiments. In this paper we introduce a self-guide method to find MBVs for unknown states using a stochastic gradient ascent algorithm (SGA), by parameterizing the corresponding Bell operators. For all the investigated systems (2-qubit, 3-qubit and 2-qutrit), this method can ascertain the MBV accurately within 100 iterations. Moreover, SGA exhibits significant superiority in efficiency, robustness and versatility compared to other possible methods.

\end{abstract}


\maketitle

\section{introduction}
In 1964, Bell proved that the conclusion of quantum theory can not be repeated by any local theory \cite{Bell}, which led to the concept of Bell nonlocality. With the development of quantum information theory, Bell nonlocality began to be considered as a potential resource in information process. As a result, a considerable amount of applications concerning Bell nonlocality have been discovered in last few decades. For instance, Bell nonlocality can be exploited to reduce communication complexity \cite{Cleve}, establish quantum cryptographic keys \cite{Ekert,Berrett,Acin}, generate random number \cite{Pironio} and assess multipartite entanglement \cite{Bancal,Brunner,Cao,Wang}. Additionally, measuring Bell nonlocality reveals the presence of entanglement in a device-independent way, which allows one to characterize an unknown source of quantum states \cite{Mayers,Bardyn,Yang,Kaniewski,Popescu}.

Bell inequalities, which lie in the heart of Bell's theorem, give us a direct access to measure Bell correlation between separated systems. In practical experiments, however, finding maximal violation of certain Bell inequalities (MBV) for unknown states tends to be difficult. Traditional methods of quantum state characterization are impractical for systems of more than a few qubits due to exponentially expensive postprocessing and data storage. In 2016, Batle proved that computing maximal Bell violation is an NP-problem \cite{Batle}, which means it is hardly possible to find MBV via ergodic method or Monte Carlo sampling. Another way to address this problem is computing MBV from full tomography results (CVT) \cite{James,Fano,Banaszek}. Unfortunately, the scaling and additional postprocessing cost make standard quantum tomography impractical for large-size quantum states being prepared today \cite{Vlastakis,Haas}. The reliability of standard quantum tomography for all system sizes is limited by sensitivity to experimental errors \cite{Braczyk,Enk}.

In this paper, we introduce a self-guide way to find MBVs for unknown states using an iterative stochastic gradient ascent algorithm (SGA) \cite{Spall}. Due to its high efficiency and robustness, this algorithm has been widely applied to fields of engineering \cite{Spall}, optics \cite{Huizhen} and quantum physics. Especially, processes including quantum tomography \cite{Chapman} and quantum control \cite{Ferrie} have been optimized using SGA. In order to indicate the feasibility of the SGA in searching MBVs for unknown states, we investigate different quantum systems including 2-qubit, 3-qubit and 2-qutrit systems. For all of these three systems, SGA converges within 100 iterations and the searched MBVs are very close to the true values.  In principle, our method can be generalized to fit systems with arbitrary dimensions and parties. The enhanced robustness of SGA against statistical noise is further demonstrated by decreasing photon number used in each iteration. Moreover, we show that SGA is superior to CVT not only when measurement errors occur, but also in some special cases that the measurement devices cannot be trusted.

\section{Algorithm and Demonstrations}
\begin{figure}
\includegraphics[width=6.6in]{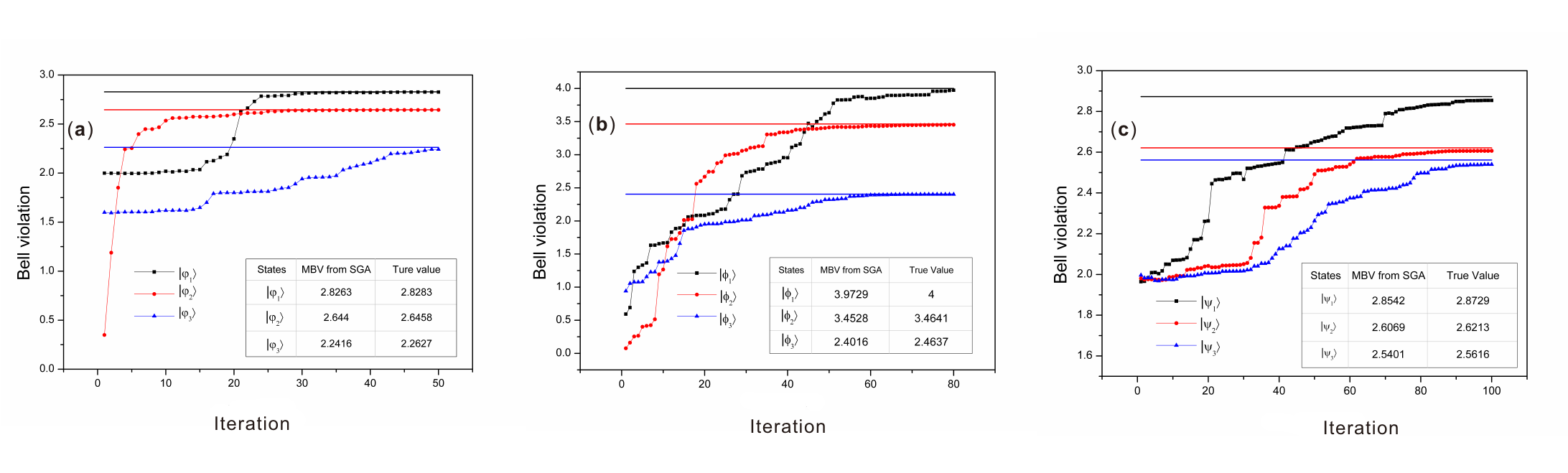}
\caption{The searching process for MBVs of (a) 2-qubit, (b) 3-qubit and (c) 2-qutrit systems when SGA is applied. For each system, different states (including pure maximally entangled state and partially mixed state) are studied and the solid lines denote the calculated true values of MBVs. The searched values of MBVs with certain interations are listed in the inset table on each figure, together with the calculated true values.}
\end{figure}

$\emph{Framework of SGA}$:
Bell operators consist of local measurements made by separated parties that share a quantum state. Since local measurements can be pinned down by several parameters (for example degree of wave plates in polarized photon scenario), the expectation value of a Bell operator $\mathcal{B}$ for certain state $|\psi\rangle$ can be determined by a group of parameters(denoted by $\Theta$), which has the form
\begin{equation}
V(\Theta)=|\langle\psi|\mathcal{B}(\Theta)|\psi\rangle|.
\end{equation}
Hence the problem of finding MBV becomes a process of maximizing a parametric function. Since this process is conducted in a high-dimensional space(e.g. 8 dimension for CHSH inequality as showed later), it is experimentally complicated to be achieved with an ergodic method. Therefore we introduce the following SGA method \cite{Spall}. As an iterative method, we generate an initial parameter $\Theta_0$ and the parameter in step $k$ is denoted by $\Theta_k$. For each step, we randomly choose a perturbation of the parameter $\Delta_k$, and measure the value of Bell operator $V(\Theta+\beta_k\Delta_k)$ and $V(\Theta-\beta_k\Delta_k)$, where $\beta_k=[b/(k+1)^t]$ controls the gradient estimation step size. Then the gradient of perturbation $\Delta_k$ is calculated by

\begin{equation}
g_k=\frac{V(\Theta_k+\beta_k\Delta_k)-V(\Theta_k-\beta_k\Delta_k)}{2\beta_k}.
\end{equation}

Then the parameter $\Theta_k$ is updated to $\Theta_{k+1}=\Theta_k+\alpha_kg_k\Delta_k$, where $\alpha_k=[a/(k+1)^s]$ controls the step size. $a,b,t$ and $s$ are algorithm parameters that can be optimized for different experimental scheme. The final value of the Bell operator is given by $V(\Theta_N)$ (i.e. searched MBV) and the number of iteration $N$ can be chosen to satisfy the experimental requirement. In the following, we will demonstrate that SGA is applicable for different size of quantum systems.

$\emph{2-qubit system}$:
Suppose that Alice and Bob share a 2-qubit system, the nonlocality can be tested with the CHSH operator \cite{Clauser} written as
\begin{equation}
\label{CHSH}
\mathcal{B}_{CHSH}=A_0B_0+A_0B_1+A_1B_0-A_1B_1,
\end{equation}
where $A_0, A_1, B_0$ and $B_1$ are local measurement operators for Alice and Bob respectively. Each operator has eigenvalues(i.e. outcomes) $\pm1$. The maximal violation of CHSH inequality given by quantum theory is $2\sqrt{2}$ while with the assumption of local theory it can only reach a maximum of 2. Notice that each local operators introduced in the CHSH inequality can be determined by two parameters. For example, we consider the common-used decomposition which gives the form $A=\vec{a}\cdot\vec{\sigma}$, where $\vec{\sigma}=(\sigma_1,\sigma_2,\sigma_3)$ are the Pauli matrices \cite{Flammia,Silva} and $\vec{a}(\theta,\phi)=(sin{\theta}cos{\phi},sin{\theta}sin{\phi},cos{\theta})$ is a three-dimensional vector denoted by the polar coordinate with identical module, giving out two parameters $\theta$ and $\phi$.

It is easy to confirm that the CHSH inequality can be determined by 8 parameters, thus we can set $\Theta$ to be an 8-dimensional vector. Setting algorithm parameters to be $a=0.2,b=0.2,t=1$ and $s=2$, we perform SGA to 3 different 2-qubit target states. As shown in Fig.1(a), for each target state, within 50 iterations, the searched MBV can be extremely close to its true value. The SGA method shows significantly high efficiency and accuracy in searching the MBV for 2-qubit state, we will further show it is likewise adequate for multi-parties and high-dimensional quantum systems.

$\emph{3-qubit system}$:
For 3-qubit system, we consider the Mermin inequality \cite{Mermin} expressed by
\begin{equation}
\label{mermin}
\mathcal{B}_{Mermin}=A_1B_0C_0+A_0B_1C_0+A_0B_0C_1-A_1B_1C_1,
\end{equation}
where $A, B$ and $C$ are local operators for three different parties similar to that of CHSH inequality, and $\Theta$ turns to be a 12-dimensional vector. When the target state is maximally entangled, i.e., a Greenberger-Horne-Zeilinger state, the MBV of Mermin inequality is $4$.

To apply our algorithm to Mermin inequality is a similar process. We keep the above algorithm parameters and choose final iteration number to be $N=80$. Fig.1(b) shows the searching process for MBVs with three different 3-qubit states, and each of the results complies well with its true value. It is worthy to note that the iteration number to achieve convergence is only $\sim1.5$ times higher than that for 2-qubit systems. In contrast, the step number of ergodic method is $\sim m^{d}$ ($m$ is the steps number for one parameter), which exponentially increases with the number of dimensions $d$. As a result, the measurement number for 3-qubit should be $\sim m^{2}$ times higher than that of 2-qubit for ergodic method.

$\emph{2-qutrit system}$:
When facing high-dimensional systems, we have to find different mathematical descriptions to parameterize the Bell inequality. In 2002, Collins \emph{et al.} \cite{Collins} developed an approach to Bell inequalities for arbitrarily $d$ dimensional two-party systems. To specify the problem, here we consider $d=3$, i.e. a 2-qutrit system shared by Alice and Bob. Then the Bell expression has the form
\begin{equation}
\begin{split}
B_{d=3}= & +[P(A_0=B_0)+P(B_0=A_1+1)+P(A_0=B_1)+P(B_1=A_0)]-  \\
         &[P(A_0=B_0-1)-P(B_0=A_1)-P(A_1=B_1-1)-P(B_1=A_0-1)].
\end{split}
\end{equation}
Where $A_0, A_1, B_0$ and $B_1$ denote the local measurements of Alice and Bob that have three possible outcomes $0, 1$, or $2$, while
\begin{equation}
P(A_a=B_b+k)=\sum_{j=0}^{2}{P(A_a=j, B_b=j+k mod 3)}
\end{equation}
denote the probabilities of the joint measurements on both sides. The maximal violation of this inequality provided by quantum physics is $4/(6\sqrt{3}-9)\approx2.873$. Seeing that the outcomes(eigenvalues) of each local operators are fixed, we can certify the operator by pin down its three orthogonal eigenvectors. For example, if eigenvectors of operator A are $|\alpha\rangle_A, |\beta\rangle_A$ and $|\gamma\rangle_A$ which give eigenvalues 0, 1 and 2 respectively, then A can be written as
\begin{equation}
A=0|\alpha\rangle_{AA}\langle{\alpha}|+1|\beta\rangle_{AA}\langle{\beta}|+2|\gamma\rangle_{AA}\langle{\gamma}|.
\end{equation}
In another word, the local operator can be expressed by a group of orthogonal basis in three-dimensional Hilbert space. According to algebra theorem, different orthogonal basis can be transformed by an unitary matrix. Thus the eigenvectors of $A$ have the form
\begin{equation}
|\alpha\rangle_A=U|0\rangle, |\beta\rangle_A=U|1\rangle, |\gamma\rangle_A=U|2\rangle,
\end{equation}
where U is the unitary matrix ($UU^{\dagger}=I$). Since an arbitrary unitary matrix can be written as $U=e^{iH}$, where $H$ is a Hermitian matrix, we can find a corresponding local measurement operator for each fixed Hermitian matrix $H$. To construct such matrices, we introduce the method of three-dimensional Gell-Mann matrices \cite{Gell}. This group of matrices can denote the Hermitian matrix $H$ in the form
\begin{equation}
H=\sum_{i=1}^{8}{\theta_i\lambda_i},
\end{equation}
where $\lambda_i(i=1-8)$ stand for the Gell-Mann matrices and $\theta_i$ denote arbitrary real parameters. Using this expression, we denote a local operator for qutrit system by 8 parameters, which results in totally 32 parameters to fix the Bell expression. Since the method of Gell-Mann matrices can be easily generalized to high-dimensional space, the above process to parameterize the Bell inequalities can be extended to arbitrary high dimensions. As for 2-qutrit system, we iterate SGA to $N=100$ for three different states. As shown in Fig.1(c), the SGA can still converge at a quantity very close to the true value for each state. Despite of the four-fold increasing in the number of parameters compared to 2-qubit system, the iteration number to achieve convergence grows only $\sim2$ times. 

\section{Test of robustness}
In contrast to theoretical computation, different kinds of errors are inevitable in real-world experiments. Therefore, it is necessary to prove that our method is robust against certain level of experiment errors. The most common and distinct errors in the measurement of polarized photon systems are statistical noise \cite{Barnett} for photon count rate and measurement imperfection caused by wave-plate uncertainty. The statistical noise can significantly degrade the measurement precision when the involved photon numbers are ultimately low. Here, in Fig. 2, we consider an assigned task to search MBV of CHSH inequality for a 2-qubit system with varying photon pair number used per measurement, denoted as $n$. The target state is selected to be the maximally entangled state $|\psi\rangle=\frac{1}{\sqrt{2}}(|01\rangle-|10\rangle)$ with a true MBV value of $\sim2.828$. For each value of $n$, we apply the algorithm with 50 iterations and repeat it ten times independently. The average of MBVs after 50 iterations are shown in Fig. 2 and we see that the SGA can still converge with a high value of MBV (about 2.76) when $n$ is reduced to 200.
\begin{figure}
\includegraphics[width=6in]{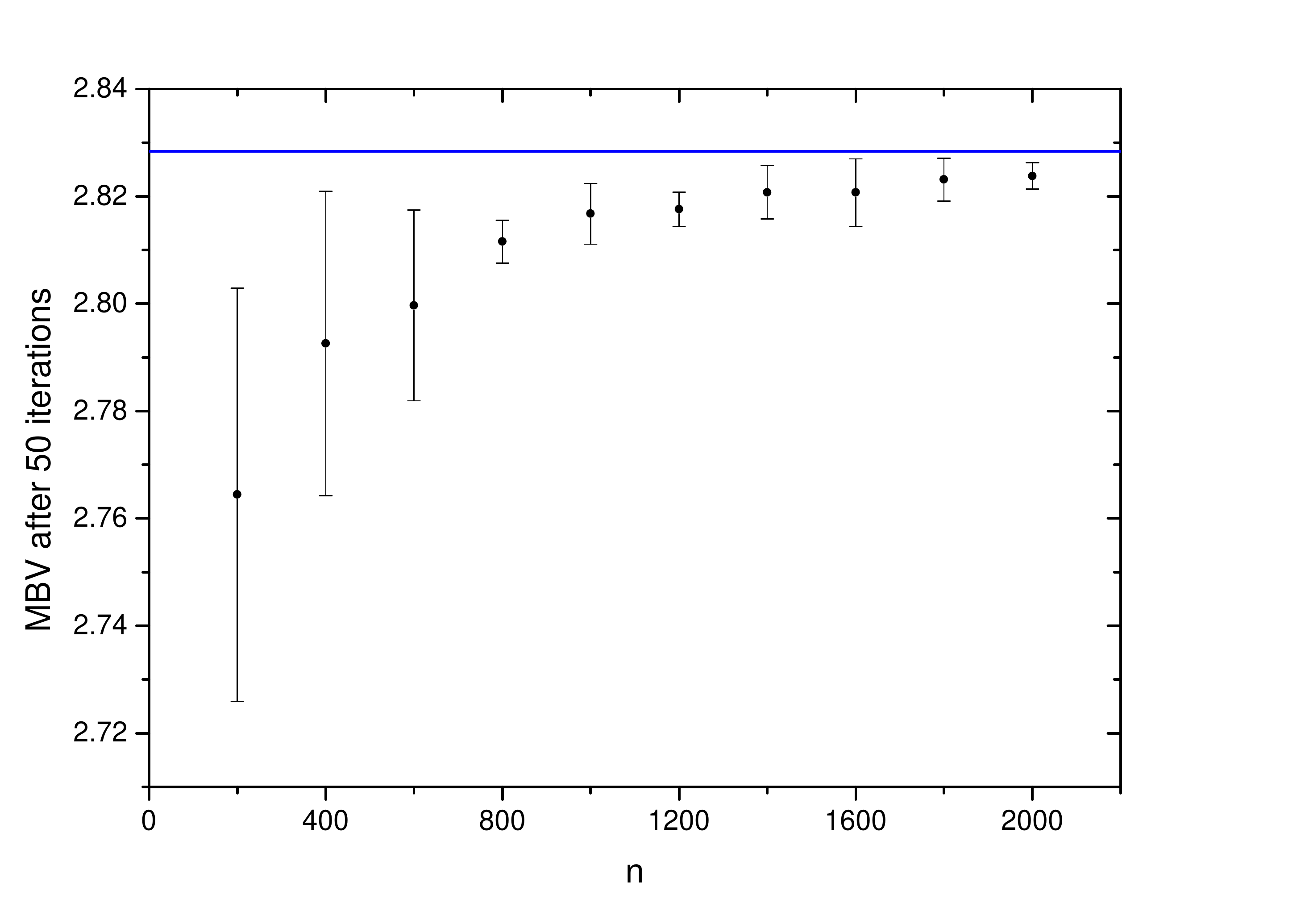}
\caption{Searched MBVs of CHSH inequality for 2-qubit singlet with different numbers of photon pairs used per measurement (i.e. different level of Poisson noise). Points denote the average of ten repetitions and the error bar gives the standard deviation. The blue solid line denotes the true value of MBS to be $\approx2.828$.}
\end{figure}

In general, the statistical noise can be largely suppressed by increasing the photon number rate or integration time. However, the technical errors are difficult to eliminated in most cases. For example, the rotating or indication errors of wave-plate always exist in Bell experiment. As a result, the robustness against technical errors maybe the decisive factor for the quality of a proposal. In order to test the usability of SGA under certain technical errors, we construct a model of which the uncertainty of wave-plates is the primary experiment errors. We engineer this uncertainty by applying random wave-plate errors under a few deviation levels. Here we still assume the target state to be 2-qubit singlet and both SGA and CVT methods are simulated to compare the robustness. Fig. 3 shows the results of these two methods in the task of searching the MBVs of CHSH inequality. In order to allow direct resource comparison, we perform the two methods using an equivalent total number of photons by running SGA with 60 iterations with unaltered algorithm parameters and repeating CVT with 60 times. Totally three levels of error are studied and for each of them SGA can outperform CVT giving a higher MBV and the results are shown in Fig. 3.

\begin{figure}
\includegraphics[width=6in]{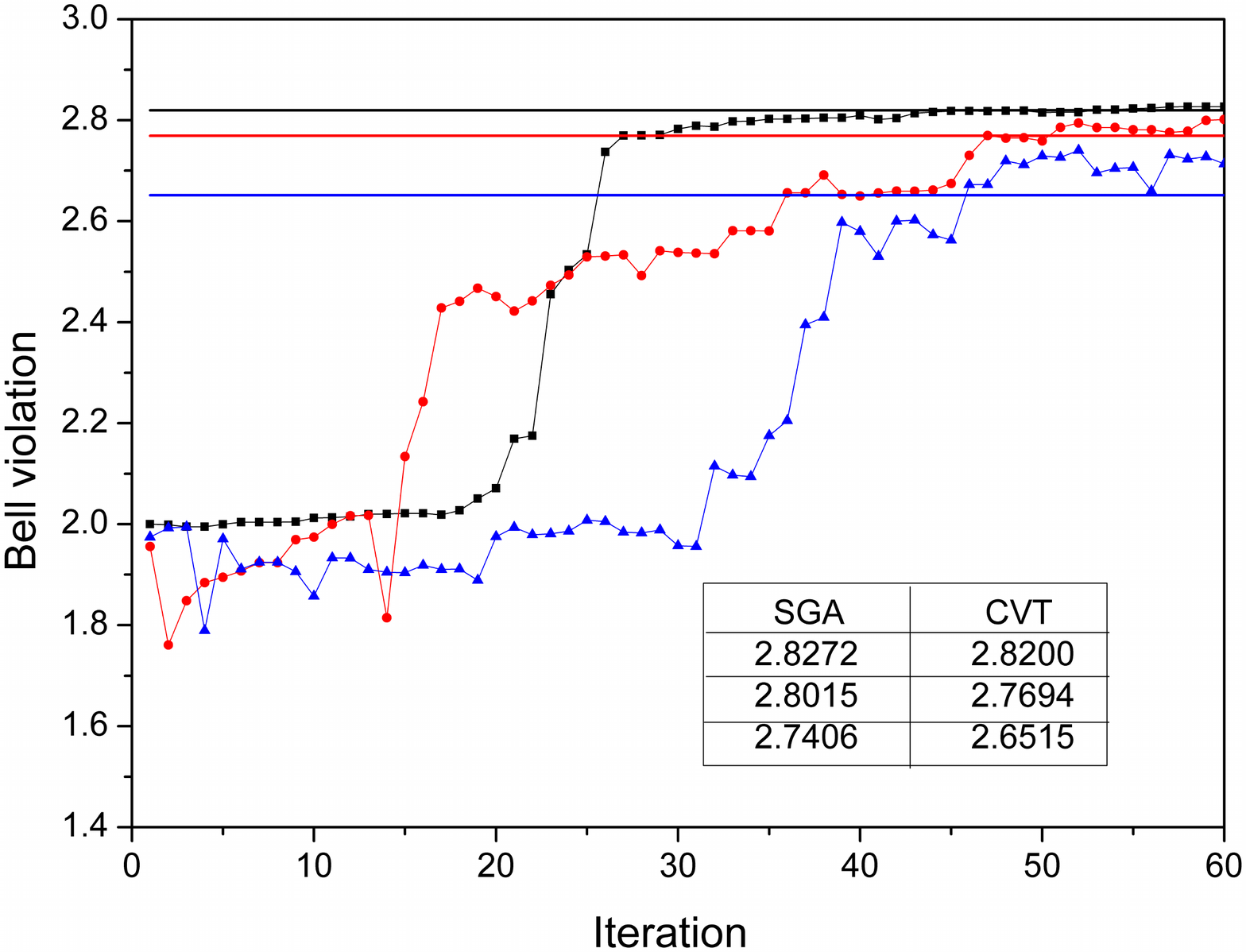}
\caption{CVT and SGA are applied to the singlet state with same level of experimental errors. The solid lines show the results of CVT while the dotted lines show the process of SGA. SGA is run for 60 iterations and CVT is repeated for 60 times. For each level of errors, SGA can outperform giving a higher MBV as shown in the inset table.}
\end{figure}

\section{With untrusted devices}

\begin{figure}
\includegraphics[width=6in]{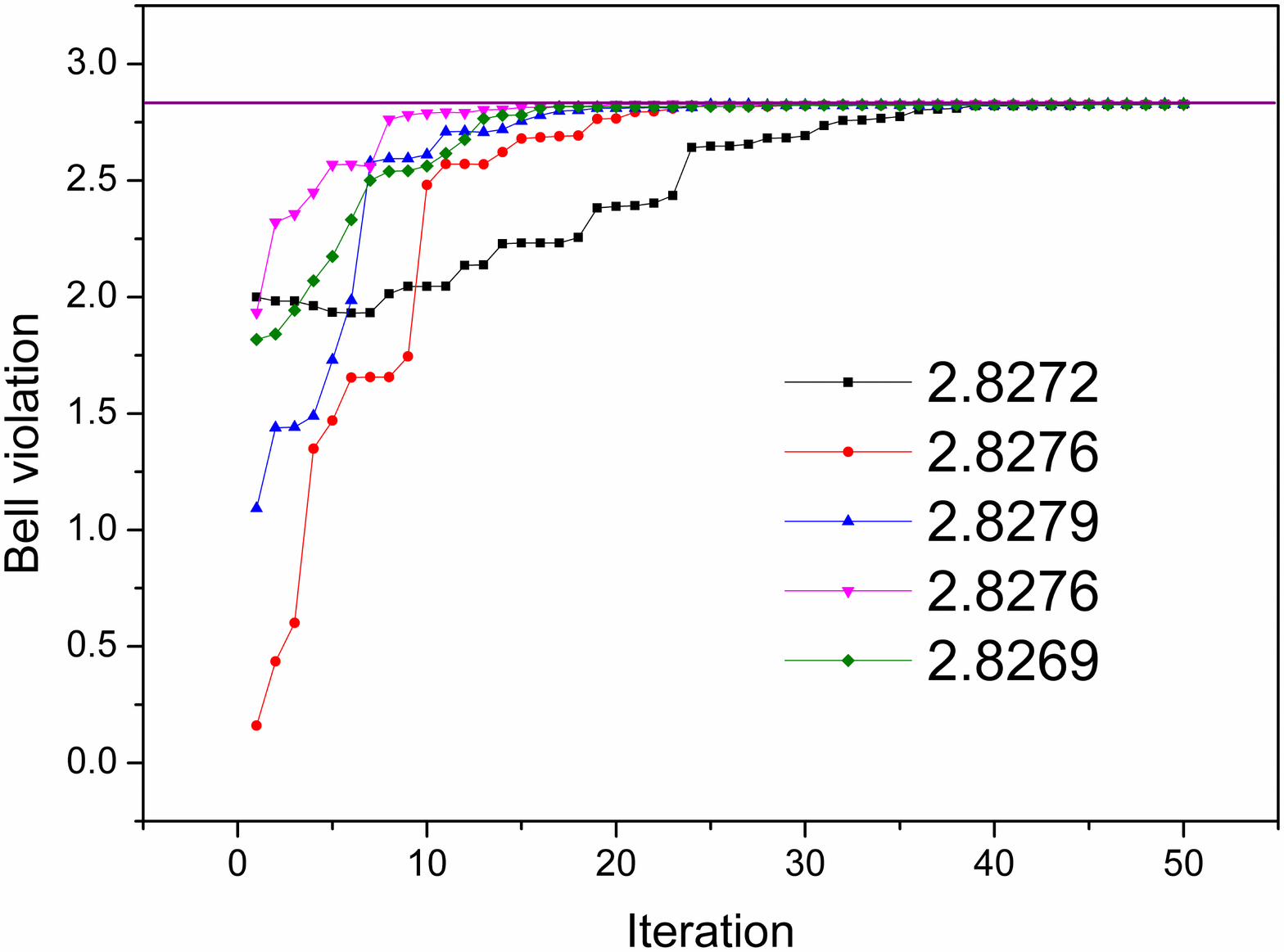}
\caption{Simulations of SGA with uncalibrated wave-plates for 2-qubit singlet. The initial estimate of operators are randomly selected and eventually they can converge at $\approx2.828$ through different routes. The purple solid line denotes the true value of 2.828 and the searched MBV from five different initial estimate of operators are shown in the inset table.}
\end{figure}

For all the above analysis, measurement devices are assumed to be perfect. Practically, however, we often do not have sufficient knowledge of the internal physical structure, or the used devices cannot be trusted. In this scenario, quantum tomography can not be implemented thus the CVT is no longer practicable for searching the MBV. However, we will show that SGA is still tolerant of some special device imperfections, i.e., the used wave-plates are not correctly, or not even, calibrated. In this case, the starting point of SGA can not be confirmed hence the current estimate of operators in each iteration are also uncertain. Yet the content of SGA method only bases on a parametrization of Bell operators and gradients are computed only by statistical outcomes, without requiring exact mathematical description of measurement operators. In another word, during the process of SGA, the updated parameter $\Theta_{k+1}$ can be completely determined by the statistical result $g_k$ and former parameter $\Theta_k$. As a result, SGA can still be feasible when the wave-plates cannot be trusted. In order to demonstrate this characristics, we randomly choose 5 initial estimate of operators and run SGA for the 2-qubit singlet with unaltered algorithm parameters. The results in Fig. 4 clearly indicate that SGA can be independent of the definite estimate of operators, and in any case it can converge at a value very close to 2.828 through different routes within 50 iterations.

\section{Discussion and Conclusion}
Finding MBV provides an approach to measure Bell nonlocality among seperated parties, which could be essential for certain tasks in quantum information process. However, traditional methods to find MBV, including ergodic by traversing all the involved operators and CVT by full state tomography, may not be applicable in many cases. Here, we propose a self-guide approach utilizing the SGA in Bell experiments. With rigorous analysis, we demonstrate the exclusive efficiency and robustness of this approach. When utilizing SGA, for all the studied systems it can converge within 100 iterations. Furthermore, in the process of searching MBV, SGA does not require data storage, computationally expensive postprocessing or maximum likelihood estimation of the target state. What is more valuable, SGA exhibits fine robustness against statistical noise and measurement errors. With the identical amount of resources, it can outperform CVT considering technical erros due to wave-plate uncertainty. Moreover, SGA is still feasible when facing some situations in which the measurement devices cannot be trusted, i.e. performing quantum tomography becomes totally impossible. To sum up, SGA is proved to be an efficient and robust method for Bell experiments where traditional methods have already become impractical.


\begin{thebibliography}{xx}
\bibitem{Bell} Bell J., Physics \textbf{1}, 195 (1964).

\bibitem{Cleve} R. Cleve and H. Buhrman, Phys. Rev. A. \textbf{56}, 1201 (1997).

\bibitem{Ekert} A.K. Ekert, Phys. Rev. Lett. \textbf{67}, 661 (1991).

\bibitem{Berrett} J. Barrett, L. Hardy, and A. Kent, Phys. Rev. Lett. \textbf{95}, 010503 (2005).

\bibitem{Acin} A. Ac\'in, N. Gisin, and L. Masanes, Phys. Rev. Lett. \textbf{97}, 120405 (2006).

\bibitem{Pironio} S. Pironio, A. Ac\'in, S. Massar, A. Boyer de la Giroday, D.N. Matsukevich, P. Maunz, S. Olmschenk, D. Hayes, L. Luo, T.A. Manning, and C. Monroe, Nature \textbf{464}, 1021 (2010).

\bibitem{Bancal} J.-D. Bancal, N. Gisin, Y.-C. Liang, and S. Pironio, Phys. Rev. Lett. \textbf{106}, 250404 (2011).

\bibitem{Brunner} N. Brunner, J. Sharam, and T. V\'ertesi, Phys. Rev. Lett. \textbf{108}, 110501 (2012).

\bibitem{Cao} D.-Y. Cao, B.-H. Liu, Z. Wang, Y.-F. Huang, C.-F. Li, and G.-C. Guo, Science Bulletin, 2015, 60(12):1128-1132.

\bibitem{Wang} Z. Wang, C. Zhang, Y.-F. Huang, B.-H. Liu, C.-F. Li, and G.-C. Guo, Science Bulletin, 2016, 61(9):714–719.

\bibitem{Bardyn} C.-E. Bardyn, T.C.H. Liew, S. Massar, M. Mckague, and V. Scarani, Phys. Rev. A. \textbf{80}, 062327 (2009).

\bibitem{Mayers} D. Mayers and A. Yao, \emph{Self testing quantum apparatus}, Quantum Inf. Comput. \textbf{4}, 273 (2004).

\bibitem{Yang} T.-H. Yang and M. Navascu\'es, Phys. Rev. A. \textbf{87}, 050102 (2013).

\bibitem{Kaniewski} J. Kaniewski, Phys. Rev. Lett. \textbf{117}, 070402 (2016).

\bibitem{Popescu} S. Popescu and D. Rohrlich, Phys. Rev. A. \textbf{169}, 411 (1992).

\bibitem{Batle} Batle, J., Ooi, C.H.R., Abdalla, S. et al. Quantum Inf Process (2016) \textbf{15}: 2649.

\bibitem{James} F.V. James, P. G. Kwiat, J. Munro, and G. White, Phys. Rev. A. \textbf{64}, 052312 (2001).

\bibitem{Fano} U. Fano, Rev. Mod. Phys. \textbf{29}, 74 (1957).

\bibitem{Banaszek} K. Banaszek, M. Cramer, and D. Gross, New J. Phys. \textbf{15}, 125020 (2013).

\bibitem{Vlastakis} B. Vlastakis, G. Kirchmair, Z. Leghtas, S. E. Nigg, L. Frunzio, S.M. Girvin, M. Mirrahimi, M.H. Devoret, and R.J. Schoelkopf, Science \textbf{342}, 607 (2013).

\bibitem{Haas} F. Haas, J. Volz, R. Gehr, J. Reichel, and J. Estve, Science \textbf{344}, 180 (2014).

\bibitem{Braczyk} A.M. Braczyk, D.H. Mahler, L.A. Rozema, A. Darabi, A.M. Steinberg, and D. F. V. James, New J. Phys. \textbf{14},085003 (2012).

\bibitem{Enk} S.J.v. Enk and R. Blume-Kohout, New J. Phys. \textbf{15}, 025024 (2013).

\bibitem{Spall} J. Spall, IEEE Trans. Autom. Control \textbf{37}, 332 (1992).

\bibitem{Huizhen} H.-Z. Yang, X.-Y. Li, Optics\&Laser technology \textbf{43} (2011) 630.

\bibitem{Chapman} R.J. Chapman, C. Ferrie,and A. Peruzzo, Phys. Rev. Lett. \textbf{117}, 040402 (2016).

\bibitem{Ferrie} C. Ferrie and O. Moussa, Phys. Rev. A. \textbf{91}, 052306 (2015).

\bibitem{Clauser} J.F. Clauser, M.A. Horne, A. Shimony, and R.A. Holt, Phys. Rev. Lett. \textbf{23}, 880 (1969).


\bibitem{Flammia} S.T. Flammia and Y.-K. Liu, Phys. Rev. Lett. \textbf{106}, 230501 (2011).

\bibitem{Silva} M.P. da Silva, O. Landon-Cardinal, and D. Poulin, Phys. Rev. Lett. \textbf{107}, 210404 (2011).

\bibitem{Mermin} N.D. Mermin, Phys. Rev. Lett. \textbf{65}, 1838 (1990).

\bibitem{Collins} D. Collins, N. Gisin, N. Linden, S. Massar, and S. Popescu, Phys. Rev. Lett. \textbf{88}, 040404 (2002).

\bibitem{GHZ} D.M. Greenberger, M.A. Horne, A. Shimony, and A. Zeilinger, Am. J.Phys. \textbf{58}, 1131 (1990).

\bibitem{Gell} M. Gell-Mann and Y. Ne��eman: The Eightfold Way, W A Benjamin 1964. Also see https://en.wikipedia.org/wiki/Gell-Mann matrices.

\bibitem{Barnett} S.M. Barnett and P.M. Radmore, \emph{Methods in Theoretical Quantum Optics} (Oxford University Press, Oxford, 1997).
\end{thebibliography}
\end{document}